# Survey On Scheduling And Radio Resources Allocation In Lte

Fayssal Bendaoud[#1], Marwen Abdennebi[*2], Fedoua Didi[#3]

[#]Laboratory of Telecommunication Tlemcen (LTT)
[*]Laboratory of Processing and Transmission of Information (L2TI)

*ABSTRACT*

*This paper focuses on an essential task of the enhanced NodeB eNodeB element in LTE architecture, the Radio Resource Manager RRM, which aims to accept or reject requests for connection to the network based on some constraints and ensuring optimal distribution of radio resources between Users Equipments UEs. Its main functionalities include Admission Control AC and Packet Scheduling PS.*

*This paper will center mainly on the PS part of the RRM task, which performs the radio resource allocation in both uplink and downlink directions. Several approaches and algorithms have been proposed in the literature to address this need (allocate resources efficiently), the diversity and multitude of algorithms is related to the factors considered for the optimal management of radio resource, specifically, the traffic type and the QoS (Quality of Service) requested by the UE.*

*In this article, an art's state of the radio resource allocation strategies and a detailed study of several scheduling algorithms proposed for LTE (uplink and downlink) are made. Therefore, we offer our evaluation and criticism.*

## 1. INTRODUCTION

Long Term Evolution (LTE) or 3.9G systems is an important technology originally designed to achieve a significant data rates (50Mbit/s in the uplink and 100Mbit/s in the downlink in a system bandwidth 20 MHz), while allowing the minimizing of the latency and providing a flexible deployment of the bandwidth. LTE offers several main benefits for the subscribers as well as to the service providers. It significantly satisfies the user's requirement by targeting the broadband mobile applications with enhanced mobility. It is designated as the successor networks 3G. It allows an efficient execution of internet services emerging in recent years. It uses the packet switching process as well as 3G networks, the difference is the using of Time Division multiplexing (TD) and Frequency Division multiplexing (FD) at the same time which is not the case of High Speed Packet Access HSPA networks, which performs only the time division multiplexing, this allows us to have a throughput gain (in spectral efficiency) concerning 40 %. [1]

Orthogonal Frequency Division Multiple Access OFDMA is the multiple access method used in the downlink direction. It combines Time Division Multiple Access TDMA and Frequency Division Multiple Access FDMA. It is derived from OFDM multiplexing, but it allows the multiple access of the radio resources shared among multiple users. The OFDMA technology divides the available bandwidth into many narrow-band subcarriers and allocates a group of subcarriers to a user based on: its requirements, current system load and system configuration, this process helps to fight the Inter Symbol Interference ISI problem or the channel frequency-selective, as well as, it allows for the same bandwidth a higher spectral efficiency (number of bits transmitted per Hertz) and it has the ability to maintain high throughput even in unfavorable environments with echoes and multipath radio waves.

For the uplink direction, Single Carrier-Frequency Division Multiple Access SC- FDMA method is used, it is a variant of OFDMA, they have the same performance (throughput, efficiency ... etc.), but SC- FDMA transmits sub bands sequentially to minimize the Peak -to- Average Power Ratio PAPR (OFDMA has a huge PAPR), this is the reason of choosing SC-FDMA in the uplink side, to deal with the limited power budget (the use of battery by the UE) to minimizing the PAPR.

An important element of the LTE architecture is the eNodeB, which has an interesting task, the RRM consisting mainly of two sub-tasks: the AC and the PS.

The AC sub-task is responsible for accepting and rejecting new requests, in fact, the decision to accept or reject a request depends on the network capacity to deliver the QoS required by the request (application) while ensuring the QoS asked by the already admitted users in the system.
The PS meanwhile, performs the radio resource allocation to the users already accepted by the AC, i.e., performing the UE-RB mapping by selecting UEs who will use the channel affecting their radios resources RBs that permit them to maximize system performance.

Several parameters can be used to evaluate the system performance such as, spectral efficiency, delay, fairness and system throughput. The variety of parameters results on the creation of multiple scheduling algorithm and strategies.

All these parameters can be summarized in one term, the consideration of flow's QoS. Trying to satisfy all these parameters is impossible, simply because the scheduling and resource allocation is an NP-hard problem, because of this; different scheduling strategies have been developed. An important parameter in the design of schedulers is the support for QoS. This forced the LTE network to differentiate between the data streams and therefore can be distinguished:

*Conversational class*: this is the most sensitive to delay; it includes video conferencing and telephony. It does not tolerate delays because it assumes that in the two ends of the connection is a human.

*Streaming class*: similar to the previous class, but it assumes that only one person is at the end of the connection, therefore, it is less demanding in terms of time and delays. Eg: video streaming
*Interactive class*: examples of this class are: web browsing, access to databases ... etc.

Unlike the previously mentioned types, data needs to be delivered in a time interval, but this type of traffic emphasizes the rate of loss of data (Packet Error Rate).

*Background class*: Also known as Best Effort class, no QoS is applied; it tolerates delays, packet loss. Examples of this class: FTP, E-mails etc. [2]

Two other parameters influence the design of scheduling algorithms in LTE uplink. The later are imposed by the SC-FDMA access method, which are: the minimization of the transmit power (up to maximize the lifetime of UEs batteries), as well, the RBs allocated to a single UE must be contiguous. This makes the radio resource allocation for LTE uplink more difficult than for the downlink.

The rest of the paper is as follows, in section 2 will be presented the mathematical modeling of the radio resource allocation problem; in section 3, a state of art of the radio resource allocation strategies and a detailed study of several scheduling algorithms proposed for LTE (uplink and downlink) is made, we will present the scheduling algorithms existing in the literature and evaluate the performance of these algorithms with some criticism in section 4, then a conclusion and perspectives will be presented in section 5.

## 2. SYSTEM MODEL

In this section we start by giving the LTE architecture, and then, we will present the mathematical formulation of the radio resource allocation problem.

### 2.1. LTE architecture

The general architecture of LTE mainly contains the Evolved Packet System EPS which includes: the Evolved Packet Core EPC and the radio part of the core network.

EPC consists of a set of control elements: Mobility Management Entity MME, Home Subscriber Server HSS, Serving Gateway and Packet-data S-GW and P-GW. The EPC is responsible for connecting with other 3GPP and non-3GPP networks. The radio part of the network is composed of eNodeB and UE. Figure 1.

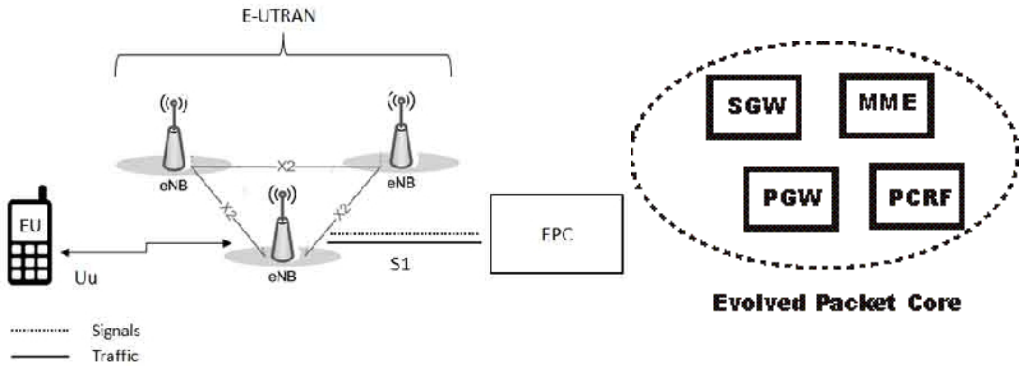

Figure 1- LTE architecture

### 2.2. The mathematical formulation of the problem

Due to the limited signaling resources, sub-carriers are often allocated in groups; that's mean, sub-carriers are grouped into Resource Blocks RBs of 12 adjacent sub-carriers with an inter-sub-carrier spacing of 15 kHz. One RB corresponds to 0.5 ms (one time slot) in the time domain, and represents 6 or 7 OFDM symbols [1]. The smallest resource unit that can be allocated to a user is a Scheduling Block (SB), which consists of two consecutive RBs, and it's the minimal quantity of radio resource that can be allocated to an UE, constituting a sub-frame time duration of 1 ms. figure 2.

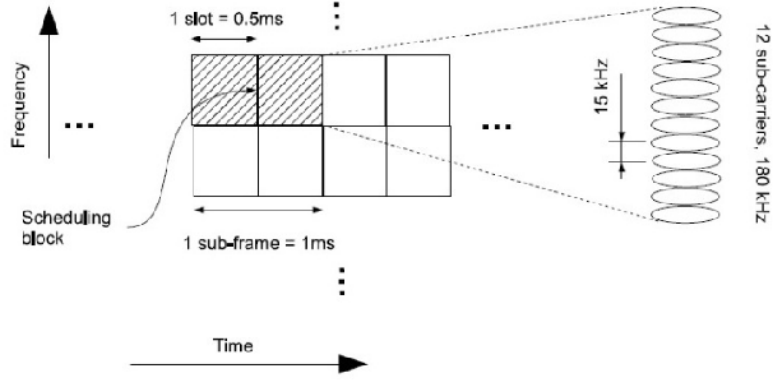

Figure 2- LTE frame structure

We consider an LTE system with N SB and K UEs, the minimum data rate required by the k-th user is $R_k$ Mbit/s. [1]

We define one SB as a set $N_s$ OFDM symbols in time domain and $N_{sc}$ sub-carriers in frequency domain. Due to control signals and other pilots, only $N_{sc}^d(s)$ of the $N_{sc}$ will be used to transmit data of the s-th OFDM symbol, with $s \in \{1,2,\ldots,N_s\}$ and $N_{sc}^d(s) \leq N_{sc}$. Assuming that

$j \in \{1,2,\ldots,J\}$, J is total number of the supported MCS (Modulation and Coding Scheme), $R_j^{(c)}$ the associated code of MCS j, $M_j$ s the constellation of the j-th MCS and $T_s$ is the OFDM symbol duration, then the achieved data rate $r^{(j)}$ by a single SB is:

$$r^{(j)} = \frac{R_j^{(c)} \log_2(M_j)}{T_s N_s} \sum_{s=1}^{N_s} N_{sc}^d(s) \qquad (1)$$

We define as CQI (Channel Quality Indicator, CQI is definite according to the modulation scheme and channel coding) of user k on the n-th SB, the CQI of user k on the N SB (all SB) is $g_k = [g_{k,1}, g_{k,2}, \ldots, g_{k,N}]$ and for all users on all SBs $G = [g_1, g_2, \ldots, g_K]$.

Each user k sends it's $g_{k,n}$ to the eNodeB to determine whose MSC must be selected by the n-th SB.

Furthermore, let $q_{k,\max(g_{k,n^*})} \in \{1,2,\ldots,J\}$ be the index of the highest-rate MCS that can be supported by user k for the n-th SB at CQI value $g_{k,n^*}$, i.e.

$$q_{k,\max}(g_{k,n^*}) = \arg\max \left( R_j^{(c)} \log_2(M_j) \big| g_{k,n^*} \right) \qquad (2)$$

The achievable throughput by user k on one sub-frame is:

$$r_k = \sum_{n=1}^{N} '_{k,n} \sum_{j=1}^{q_{k,\max}(g_{k,n^*})} b_{kj} r^{(j)} \qquad (3)$$

Where:

$\alpha_{k,n}$=1 if the n-th SB is allocated to k-th user, and $\alpha_{k',n}$ = 0 for all k' ≠ k (one SB is assigned to one and a single user).

$b_{k,j}$ Is the MCS selected by the user k on all SBs allocated to it, $b_{k,j}$=1 means that the j-th MSC is chosen by the user k.

Therefore, the radio resource allocation problem can be reported to the throughput maximization for all users under the following constraints:

$$\max_{\alpha_{k,n}, b_{k,j}} \sum_{k=1}^{K} r_k \quad (4)$$

Constraint to:
$$r_k \geq R_k \quad \forall k \quad (5)$$

$$\alpha_{k,n} = 1, \; \alpha_{k',n} = 0 \quad \forall k' \neq k \quad (6)$$

$$\sum_{j=1}^{q_{k,\max}(g_{k,n'})} b_{k,j} = 1 \quad (7)$$

(4), represents the objective function that is designed to maximize the data rate. (5), means the constraint that aims to guarantee the minimal data rate for each user. (6), is constraint assuring that one SB is assigned to one and a single user. (7), all SBs owed to a user employ the same MSC (it is an LTE networks constraint).

In literature, it is proven that the problem (4) is an NP-hard one, after that several authors have proposed their algorithms aimed solving it.

## 3. SCHEDULING IN LTE

One of the main features in LTE systems is the multi-user scheduling because it is in charge of satisfying the QoS of all active users.

In this section, we present an art's state on existing scheduling algorithms for both directions downlink and uplink. These algorithms are based on the mathematical formulations mentioned above, try performing the efficiency radio resource allocation to the system's users.

### 3.1. Downlink scheduling algorithms

The central objective of LTE scheduling is to satisfy Quality of Service QoS requirements of all users by trying to reach an optimal trade-off between efficiency and fairness. This goal is very challenging, especially in the presence of real-time multimedia applications, which are characterized by strict constraints on packet delay and jitter.

The radio resource allocation algorithms, aims to improve system performance by increasing the network spectral efficiency and fairness. It is therefore essential to find a compromise between efficiency (rate increasing) and fairness among users. Several families and categories of algorithm exist in literature; each family usually contains a set of algorithms that have a common characteristics.

### 3.1.1. Opportunistic algorithms

Opportunistic scheduling considers user where queues are continuously backlogged (this full buffer setting is typically used to model elastic or best effort flows). The main objective of this type of algorithm is to maximize the overall system throughput. Several algorithms use this approach such as: PF (Proportional Fair), EXP-PF (Proportional Fair Exponential) and the M-LWDF (Maximum Largest Weighted Delay First) scheduler is an opportunistic scheduler and also a delay based one, so we will describe it later in the delay based algorithms section.

- Proportional Fair PF

It is known that the spirit of the 4G networks is the utilization of multimedia flows, which have an important dependence with delay because they are performed in real time. Unfortunately, the PF algorithm does not consider the packet delay and the Head of Line HoL during the resource allocation process. [10]

On the other hand, the PF algorithm is a very appropriate scheduling option for non real-time traffic; the purpose is to maximize the overall throughput of system by increasing the spectral efficiency of all users together while trying to ensure fairness between users, the objective function representing the PF algorithm is:

$$a = \frac{d_i(t)}{d_i^-} \tag{8}$$

$d_i(t)$ : Data rate corresponding to CQI of user i.
$d_i^-$ : The maximum data rate supported by the RB.

- EXP-PF

This is an improvement of the PF algorithm that supports real-time streams (multimedia); in fact, it prioritizes the real time flows with respect to the other ones [14][15]. The user k is designated for scheduling according to the following relationship:

$$k = \max_i a_i \frac{d_i(t)}{d_i^-} \exp\left(\frac{a_i W_i(t) - X}{1 + \sqrt{X}}\right) \tag{10}$$

$$X = \frac{1}{N} \sum_i a_i W_i(t) \tag{11}$$

$W_i(t)$ : The HOL packet delay.
$a_i$ : Strictly positive parameter.

For the non elastic flows (best effort flows), the HOL packets delay is similar for all users (do not differ a lot), the exponential term is closed to 1, and the EXP-PF perform as the PF algorithm.

### 3.1.2. Fair algorithms

We must know that equity or fairness does not mean equality. The main objective of this category is to reach fairness and equity between users. Generally, these algorithms have an insufficiency in term of spectral efficiency. Several works have treated the fairness between users like, Round Robin, Max-Min and game theory based algorithms.

- Round Robin RR

The famous RR algorithm is largely used in the radio resource allocation because of its simplicity and low-complexity; it is dedicated to treat the fairness problem between users. So the algorithm allocates the same amount of resource sharing the time. This strategy lack in spectral efficiency and throughput performance, because the algorithm does not consider the reported SINR values when performing the allocation process.

- Max-Min Fair

The algorithm allocates resources among users successively in order to increase the data rate of each user. Once the user assigns resources required to achieve its required data rate, the algorithm select another user for scheduling. The algorithm stops when satisfying all user or all resources were allocated.

### 3.1.3. Throughput Based Algorithms

This type of algorithm tries to maximize the objective function that represents the data rate, this approach treats the real time flows and non real-time, the resource allocation depends on the size of the queue of each user. EXP Rule, Max -Weight are examples of this category.

- EXP Rule

Its main objective is to serves high data rate requirements; [16] it is represented by the following relationship:

$$k = \max_{i} \ \exp\left(\frac{a_i \ W_i(t)}{1 + \sqrt{X}}\right) \frac{d_i(t)}{d_i^-} \quad (12)$$

$$a_i = \frac{6}{\alpha_i} \quad (13)$$

$\alpha_i$ : is the maximum target delay of the i-th flow.

- *Max-Weight*

This algorithm use the packet delay criterion in the scheduling decision, its mathematical formulation is:

$$a = W_i(t) \frac{d_i(t)}{d_i^-} \quad (14)$$

Where, all parameters are explained above.

### 3.1.4. QoS Based Algorithms

This type of algorithm focuses -on the spectral efficiency of real time or non-elastic flows (video and VoIP), indeed, it tries to maximize the objective function that represents the data rate, this approach treats the real time flows and for the non real-time flows, it considers that they do not deserve any priority. The radio resource allocation process depends on the size of the queue of each user.

### 3.1.5. Delay Based Algorithms

HOL and delay of packet flows are the fundamental parameters of this kind of schedulers. This type treats the non elastic flows, when a packet exceeds its HOL, it will be removed from the queue. M-LWDF is a delay based algorithm and in the same time an opportunistic one.

- M-LWDF

M-LWDF supports multiple data flows with different QoS requirements. This algorithm is dedicated for real time services. Its decisions are based on the HOL and packets delay values. Unfortunately, it is not a suitable choice when performing non real time flows because packets delay does not have a significant role [12]. The scheduling formulation is:

$$k = \max_i a_i \frac{d_i(t)}{d_i^-} W_i(t) \qquad (15)$$
[

This is substantially the same formula of the EXP-PF algorithm, except that $a_i = -\log(p_i)T_i$, with

$p_i$ : The probability that the delay is not respected.
$T_i$ : Delay tolerated by user i.

### 3.1.6. Multiclass Algorithm

This approach considers flows classes when the treatment is different for each class RT and NRT. This type algorithm favors real time flows compared to not real time ones, which makes it the most suitable and more effective scheduling in LTE networks, but equity is not really considered.

### 3.2. Uplink Radio Resource Allocation

Unlike downlink scheduling, uplink scheduling side is much more complicated for several reasons, first, the UE sends the data to the eNodeB and we know very well that the UE has a limited energy source; secondly, it is very difficult to predict the number of radio resources that the UE needs to exchange data with the base station. Depending on the objective function considered and the traffic classes that pass over the radio channels, we have three different categories of schedulers: those dealing with best-effort flows (best effort schedulers), whose take into account the QoS and those optimizing the power transmission. In this section we will try to turn on the main families of algorithms for resource allocation in LTE uplink.

### 3.2.1. Paradigms of matrix construction

For the LTE uplink radio resource allocation, the scheduler has in input a UE-RB association matrix in order to give the best results that improve the system performances.

In the matrix creation process, literately speaking, there exist two major patterns or paradigms Channel Dependent CD and Proportional Fair PF.

In the matrix creating process, the CD paradigm considers the channel state information (CSI), so, users whom have larger CSI values will have the opportunity to allocate more resources, this approach reaches high throughput values but suffers from starvation problem. Meanwhile, the PF paradigm use the ratio of CSI and data rate of each user, so fairness is proportional on CSI values. This approach achieved good throughputs and at the same time, it solves the starvation problem. [4]

### 3.2.2. LTE Uplink system modeling

The uplink scheduling algorithms take in input a matrix with K rows (number of active UEs) and M columns (number of RBs). $M_{i,m}$ is a matrix value associated to the couple $(UE_i - RB_m)$. Following the paradigm used, this value correspond to the CSI for each RB on each UE (channel dependent), or the ratio between CSI and he data date of each user. [4][13]

|  | $RB_1$ | $RB_2$ | ... | $RB_M$ |
|---|---|---|---|---|
| $UE_1$ | $M_{1,1}$ | $M_{1,2}$ | ... | $M_{1,M}$ |
| $UE_2$ | $M_{2,1}$ | $M_{2,2}$ | ... | $M_{2,M}$ |
| ⋮ | ⋮ |  | ... |  |
| $UE_K$ | $M_{K,1}$ | $M_{K,2}$ | ... | $M_{K,M}$ |

Figure-3 UE-RB matrix association

As we said, the value in the matrix represents the association between UEs and RBs, thereafter, these values will be used by the scheduler.

### 3.2.3. Uplink scheduling algorithms

In this sub-section, we will give an art's state of the well known scheduling algorithms families for the LTE uplink side.

- Legacy schedulers

This family contains the famous classical algorithm, the Round Robin algorithm; it is also called the base scheduler's family, the RR algorithm has been used in many old systems.

The Round Robin algorithm principle is to divide the available RBs into groups of RBs according to $\left\lfloor \frac{|N_{RB}|}{|N_{UE}|} \right\rfloor$. Then, distribute the formed groups among available UEs.

- Best effort schedulers

The main objective of this category is to maximize the utilization of the radio resource and the equity in the system. It doesn't mean that this category treat only best effort flows, best effort schedulers means it is a greedy algorithm that try to do the better that it can.

As we have already said, each algorithm has an objective function to optimize, this type of algorithm uses the PF metric.

Several algorithms have been proposed in this family, we noted that the greedy algorithms are very suitable for this kind of traffic.

The principle of greedy algorithm is that the RBs are grouped into RCs, with each RC containing a set of contiguous RBs. After that each RC gets allocated to the UE having the highest metric in the matrix, the RC and UE will be removed from the available RC list and UE schedulable list. The algorithm aims to maximize the fairness in resource allocation among UEs.

This algorithm uses the PF paradigm and tries to maximize the following objective function

$$U = \sum_{u \in \overline{U}} \ln R(u) \qquad (16)$$

R(u) represents the data rate at instant t. The using of the logarithmic function is to have the proportional fairness.

After that, authors in [5] have proposed three algorithms, First Maximum Expansion (FME), Recursive Maximum Expansion (RME) and Minimum Area Difference. They belong to the same category, so they use the same objective function, but they differ in the manner that the resources were allocated.

For the FME, the algorithm starts with searching the UE having the highest metric value, once found, it expands the allocation process in the left or in the right (it compares the value of RB in the left with right value for the same UE and chooses the highest), until the algorithm finds no more RB whose having highest metric for the same user selected above.

In the other hand, The RME scheduler starts similarly to FME (it searches for the couple (UE-RB) having the highest metric value), then it expands the allocation process both in the left and the right until there will be no more users whose maximum metrics belong to the same user.
The MAD algorithm is a search-tree based; its problem is having a higher computational complexity.

It has been proven that RME has higher performance compared to FME in term of spectral efficiency. So, after that, RME has been explored in [17][18], the authors proposed two variants of RME, the Improved RME (IRME), and the Improved Tree-Based RME (ITRME).The results show an improvement in spectral efficiency by 15%.

- QoS based algorithms

Two important elements must be taken into account by this scheduler's family, the maximum delay and the throughput. Also the algorithm must offer the required QoS parameters for each user regarding to the already served users.

The Proportional Fair with Guaranteed Bit Rate (PFGBR) is a QoS based algorithm, From its name, we identify two metrics, PF and GBR, the PF metric is used to schedule the UEs with non GBR flows and for those with GBR flows , the algorithm changes the metric in order to differentiate the EU (giving priorities for UEs handling GBR streams). This algorithm has two objectives, maximizing the fairness of non GBR flows and preserves the QoS of GBR. The objective function is as follows. [6]

$$M(u,c) = \begin{cases} \exp\left(\alpha \cdot (R_{GBR} - R^-(u))\right) \cdot \frac{R^*(u,c)}{R(u)} & u \in U_{GBR} \\ \frac{R^*(u,c)}{R(u)} & u \in U_{non-GBR} \end{cases} \quad (17)$$

$R^-(u)$ : Average throughput of user u at TTI t
$R(u,c)$ : Estimated throughput of user u at resource chunk c at TTI t. Resource Chunk RC is a set of continues RBs.

This algorithm performs very well with the UEs having QoS requirement ant treats the starvation problem of UEs handling best effort traffic.

The authors in [7] have proposed two algorithms, they use a combined utility based metric with guaranteed bit rate and delay provisioning. The objective function used is defined as follows:

$$\max \sum_{u \in U} \sum_{r \in RB} x_{u,r} \cdot f_r \quad (18)$$

$x_{u,r} := 1$ if the RB r is allocated to the UE u.

$f_r$ is defining as :

$$f_r = \frac{R_u * D_i^{max}}{R_i^{min} * D_i^{avg}} \tag{19}$$

$R_u$ : acheivable throughput.
$R_i^{min}$ : minimal throughput of the i-th service class.
$D_i^{max}$ : maximum dealy of i-th service class.
$D_i^{avg}$ : average dealy of i-th service class.

The first one, named Single Channel Scheduling Algorithm SC-SA assigns one RB to each UE at a given TTI. In case that number of active users is lesser than the number of RBs, the algorithm distributes the RBs proportionally between users according to $\frac{N_{RB}}{N_u}$. Otherwise i.e. if the number of schedulable users is higher than the total number of available RBs, it assigns RBs to users experiencing the poorest conditions (eg, users that the maximum delay is almost reached). The main objective of this algorithm is to allocate resources to UEs with severe QoS constraints.
The second is called Multiple Channel Scheduling Algorithm (MC-SA). It is similar to the first one; the main difference is the possibility to allocate more than one RB to the users that are not meeting the throughput target. These algorithms have the same behavior in case the number of UEs is smaller than the number of available RBs in the system. In the case where the number UE is higher than that of available RBs, it allocates the $\left\lceil \frac{R_i^{min}}{R_u} \right\rceil$ taking into account the (19) equation.

It starts with bad conditions one; it first looks for all the RB that maximizes data rate and then looks at the left and right of this RB to the allocation of remaining n RBs.

- Power-Optimizing schedulers

The main purpose of this class of algorithms is to minimize the transmitted signal power trying to extend the duration activity of UE. In fact, it coincides with the objective of using SC-FDMA method. Schedulers of this family usually have some QoS treatments, so they perform some decisions to reduce the transmitted power till maintaining the minimal QoS requirements. This approach was not really too addressed by researchers, therefore, few algorithms appear in the literature. Such as [8][9]

## 4. DISCUSSION ET EVALUATION

The PF scheduler is often used in 3G networks, as the rate of this type of network is limited. For beyond 3G networks, a key factor must be taken in the mind, the maximum delay of multimedia flows that represents the type of the most important traffic in the B3G networks. Unfortunately this factor is not considered by this algorithm, consequently, for the non-real time flows it works fine but for the real time traffic it is not a smart choice.

Concerning EXP-PF, the parameters $W_i(t)$ and $a_i$ define the required QoS level by the flow. These parameters try to give more importance to applications with a higher level of QoS. When the exponential part of the formula is equal to one, the EXP-PF algorithm performs like the PF algorithm. This scenario is possible if the flows have almost the same delay for different users.
The RR algorithms does not take into account the QoS, because the flow does not have same needs (VoIP, Streaming etc.), also allocate the same amount of resources is not really fair, because users have not necessarily the same channel conditions, nor same types of flows etc. Beyond 3G networks, LTE specifically focuses on QoS of real time flows, so, using RR is really not the good choice.

Trying to satisfy all users in the MMF algorithm, gives the advantage to users with low requirements, so they will often served. In the other side, Users who require more resources are penalized. This approach does not take into account the multiuser diversity and the fact that streams have different QoS requirements i.e. fairness does not represent equality. To summarize, this algorithm is really not the right choice for scheduling in LTE.

As a conclusion, the radio resource allocation is feasible (several algorithms and approaches exist), but the diversity of flows (QoS) and radio conditions affect the performance of the algorithms. The resource allocation is an NP-hard problem, since the algorithm tries to maximize and/or minimize several parameters simultaneously. For this reason, each approach or algorithm tries to optimize the maximum parameters that could.

Concerning uplink side, it is much more complicated for the two SC-FDMA constraints added, the RBs allocated to a single user must be continuous, and the signal power transmitted constraint. Algorithms dealing with QoS are best suited and most respondents because they treat the most important factor in LTE networks, which is the QoS flows. But also the best effort schedulers have a good performance and there are most used because of their low complexity.

## 5. CONCLUSION

The Radio Resource Allocation is made in the eNB by the PS, this task is too complex, as it requires taking into account several factors at the same time, plus it must be immediate (real time). The objective of this paper is to present a state of the art on the radio resource allocation in LTE. In this work, we tried to go about the existed approaches in the literature in both downlink and uplink directions, we also mentioned some algorithms, we have shown the advantages and disadvantages of each category and algorithms, thereafter it would be wiser to focus on one type of traffic, trying to improve performance, and without doubt it will be the multimedia flows.